# Folding of the apolipoprotein A1 driven by the salt concentration as a possible mechanism to improve cholesterol trapping

M. A. Balderas Altamirano[*,1], A. Gama Goicochea[1] and E. Pérez[1]

**Abstract**

The folding of the cholesterol − trapping apolipoprotein A1 in aqueous solution at increasing ionic strength is studied using atomically detailed molecular dynamics simulations. We calculate various structural properties to characterize the conformation of the protein, such as the radius of gyration, the radial distribution function and the end to end distance. Additionally we report information using tools specifically tailored for the characterization of proteins, such as the mean smallest distance matrix and the Ramachandran plot. We find that two qualitatively different configurations of this protein are preferred: one where the protein is extended, and one where it forms loops or closed structures. It is argued that the latter promote the association of the protein with cholesterol and other fatty acids.

## 1 Introduction

The apolipoprotein A1 (APOA1) is the main component of high-density lipoproteins and has an important role in lipid metabolism. APOA1 is found in the human blood stream and promotes fatty acid efflux, including cholesterol, from tissues to the liver for excretion. Its study is important because of its protective effect against atherosclerosis (Breslow, 1996) and its function as a cholesterol transport from white blood cells within artery walls (Glomset, 1968). APOA1 contains a single polypeptide chain of 243 amino acid residues (Brewer et al., 1978) with 11 or 22 regularly repeating residues in the sequence (McLachlan, 1977). These multiple repeating units were proposed to form amphipathic helices with distinct hydrophilic and hydrophobic faces (Segrest et al., 1974). The fat transport is then expected to be modulated by the hydrophilic and hydrophobic residues along APOA1, but also by physiological conditions like pH and ionic strength in the blood or physiological fluid.

[1] Instituto de Física, Universidad Autónoma de San Luis Potosí, Álvaro Obregón 64, San Luis Potosí 78000, San Luis Potosí, Mexico. [*]Corresponding author: M. A. Balderas Altamirano. Electronic mail: miguelangel.balderas@gmail.com .

Computer simulations have come to play an increasingly important role in the understanding of protein folding, protein – protein interactions and the surface activity of proteins, to name but a few. Some of the advantages of atomistic simulations are the complete control over the physicochemical variables of the problem and the essentially exact solution of the equations of motion that govern the nature of the system, not to mention the vivid representations of the spatial conformations of molecules and their evolution with time. In this work we focus mainly on predicting how APOA1 folds as the salt content is increased, using atomistically detailed computer simulations. In particular, we study the contraction followed by re expansion of some residues of APOA1 induced by the increasing concentration of monovalent salt ions. This phenomenon has been found to occur in other polyelectrolytes in aqueous solution using a variety of methods [Hsiao, 2006; Feng, 2009; Alarcón, 2013; Pollard, 2013; Frank, 2000]. In this context, APOA1 is a much more complex system because there are positive, negative and neutral sequences along its structure, as well as hydrophilic and hydrophobic residues. We expect that the spatial conformation is a crucial factor in determining the ability of APOA1 to capture cholesterol and other fatty acids, which is the functionality we would like to optimize so that a mechanism can be proposed to help design drugs and treatments that improve the quality of life.

## 2 Model and Methodology

We start by taking the fragment of APOA1 known as 1GW3 from the Protein Data Bank [PDB], which has 142 – 187 aminoacid residues from the complete APOA1. It is computationally very demanding to model the entire protein, and besides only a fraction of its residues are responsible for the folding we are interested in modeling. Only a 20% of the original protein we take account. Therefore we work only with the above mentioned fragment and set up molecular dynamics simulations with it and a varying NaCl concentrations, ranging from 0.01 M up to 2.0 M. For the interatomic interactions we used the Lennard – Jones model [Allen, 1987], while for the electrostatic interactions we used the so called Particle Mesh Ewald (PME) method [Darden, 1993]. To conserve the bonds between the atoms that make up the protein we used LINCS (Linear Constraint Solver, [Hess, 1997]). The force field parameters for the protein where taken from OPLS (Optimized Potential for Liquid Simulations) [Jorgensen, 1988], and the water model used was SPCE [Berendsen, 1987]. Then all interactions are solved using GROMACS 4.6.4 [van der Spoel, 2005], where the interactions are calculated at every time step using the Verlet scheme [Pall, 2013] with a grid scheme for GPU's, which allow us to perform large simulations. The cut off distance for the Lennard – Jones and electrostatic interactions was equal to 1.0 nm. The leap – frog algorithm [Snyman, 2000] was used for the calculation of the atoms positions and velocities. The energy minimization was performed using the steepest descends method [Chaichian, 2001]. The simulations were carried out under the thermodynamic conditions known as the canonical ensemble, where the number of particles, volume of the simulation box, and temperature (*NVT*) are constant; the latter was fixed at 300 K using the V-rescale method [Buss, 2007]. To bring the system to equilibrium we ran the simulations for up to 200 ps, after which we switched to the *NPT* ensemble (where the pressure is held constant, in addition to the number of particles and the temperature) to fix the pressure at 1 bar, using the Parrinello – Rahman method [Parrinello, 1981], again for 200 ps. Once the pressure of the system is equilibrated under these conditions, we ran the simulations for an additional 5 ns for the equilibrium phase and another 5 ns for the production phase, during which we performed the calculation of the properties of interest. The time step for the integration of the equation of motion was equal to 2 fs. We worked on a cubic simulation box with lateral size equal to 10 nm.

## 3 Results and Discussion

Since we are interested in determining how APOA1 folds, we performed simulations to calculate its radius of gyration ($R_g$) at each of the salt concentrations we modeled. $R_g$ is calculated from the center of mass of the molecule, as is done in polymer science [Grosberg, 1994]. Figure 1 shows the values of $R_g$ at every one of the NaCl concentration; we have also included snapshots of the spatial configuration of the protein at certain salt concentration where the folding changes intermittently from loops (or "closed" conformations) to an extended (or "open") conformation. At the top part of Fig. 1 we include snapshots of the open conformations and in the bottom part the close conformations. The arrows at the top of Fig.1 indicate the 0.5, 0.9, 1.1, 1.4 and 1.8 M concentrations, where the open conformations take place.

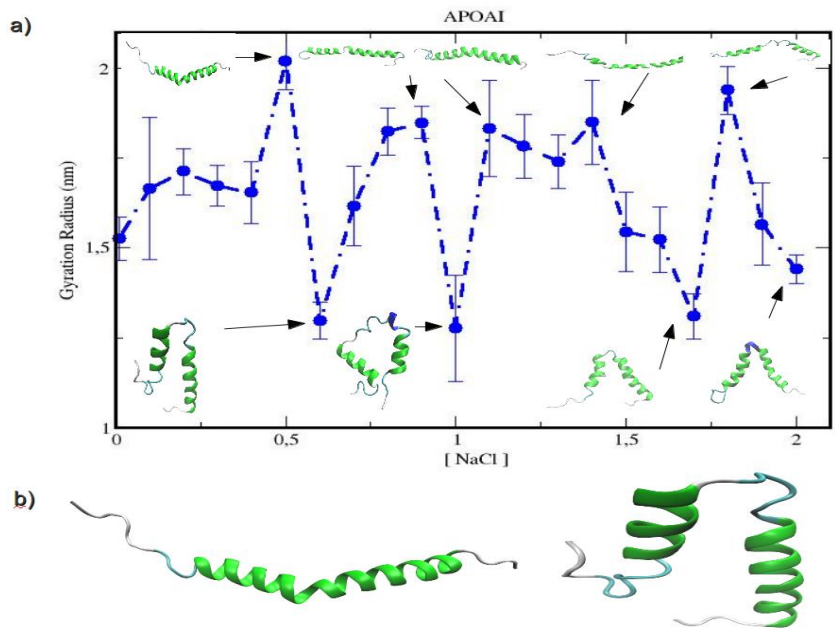

**Figure 1**. a) Gyration radius of APOA1 as a function of NaCl concentrations. The error bars represent the standard deviation of the data point averaged over 5 ns. The snapshots shown at the top and bottom of the data correspond to the open and close conformation of APOA1, respectively. The arrows identify each conformation with its salt concentration. The line is only a guide for the eye. b) Typical APOA1 open (0.5M) and closed (0.6M) conformations.

The arrows shown at the bottom of Fig. 1 represent the concentrations 0.6, 1.0, 1.7 and 2.0 M, which correspond to close conformations. Behavior qualitatively similar to this one has been obtained by Alarcón and coworkers [Alarcón, 2013] on biopolymers, by Hsiao and Luijten [Hsiao, 2006] for polyelectrolytes of multivalent ions, and in experiments [Kozer, 2007;

Dawson, 2007; Käs, 1996]. The formation of closed configurations is the result of the electrostatic charge inversion mechanism [Nguyen, 2000], while the open configurations are attributed to electrostatic repulsion between adjacent monomers [Wong, 2010]. Trapping of cholesterol and other fatty acids becomes favorable when closed configurations are formed because these molecules are closer to APOA1, therefore the formation of hydrogen bonds between them is promoted. Under these circumstances this protein would have an increased functionality. To test the stability of the open and close configurations seen in Fig. 1 we monitored their evolution in time, which we show in Fig. 2. The curves seen on the top of the figure correspond to open structures, while those at the bottom are for the closed conformations.

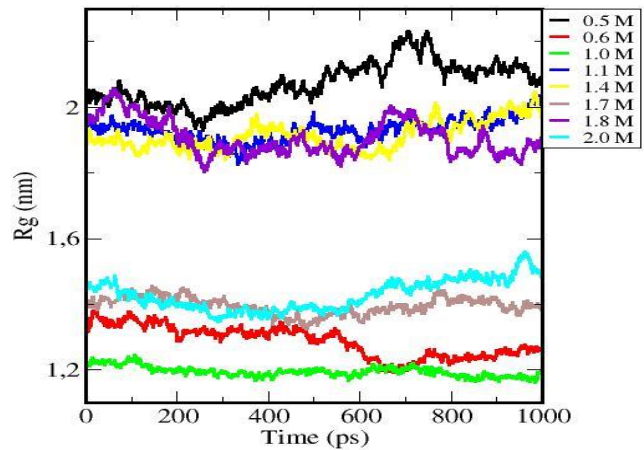

**Figure 2**. Evolution in time of the radius of gyration ($R_g$) of APOA1 at various NaCl concentrations, shown in the right panel of the figure. The top curves correspond to the so called open structures; those at the bottom correspond to closed configurations.

Inspection of Fig. 2 shows that most conformations are stable throughout the production phase of the simulation, which indicates they correspond to situations of thermodynamic equilibrium of the system, at those salt concentrations. Although in some cases one observes an increase (as in, for example, the cyan and black curves) or a decrease (red and purple curves) of $R_g$ with time, those fluctuations amount to no more than a few percent change of the averaged $R_g$.

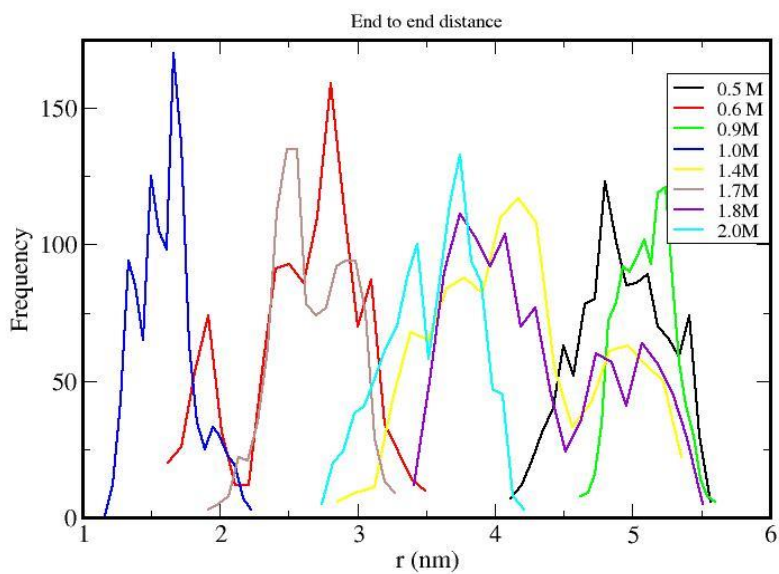

**Figure 3**. Histograms of the distance between one end of APOA1 and the other, for various concentrations of salt ions. Open conformations correspond to salt concentrations equal to 0.5, 0.9, 1.4 and 1.8 M, and the rest belong to closed configurations.

It should be stressed that both open and closed configurations have a distribution of values of the radius of gyration rather than a single value. To prove this fact we calculate the end to end distance distribution, and we report it in Fig. 3. The end to end distance is the distance from the first to the last residues in the protein. The curves seen in Fig. 3 represent histograms of the distribution of end to end distances found at various salt concentrations. The so called open configurations have larger end to end distances, as is to be expected, corresponding to the concentrations 0.5 M (black line), 0.9 M (green), 1.4 M (yellow) and 1.8 M (purple). The closed conformations have smaller end to end distances, as prove the lines for 0.6 M (red line in Fig. 3), 1.0 M (blue), 1.7 M (gray), and 2.0 M (cyan). Most of this information was known at this point, particularly from the analysis of Figs. 1 and 2. However, the added value of Fig. 3 is that it provides a quantitative estimate of the average distance between the ends of APOA1 when it is found in what we call an open configuration (the largest being ~ 5.3 nm) and in a closed configuration as well (the smallest is about 1.6 nm).

The structural characteristics of APOA1 are further analyzed through the so called "mean smallest distance matrix" (msdm), which is obtained from the averaged distance between each residue and all other residues in the protein. In Fig. 4, the *x* and *y* axes indicate the index that identifies each residue of the protein, and each data point in Fig. 4 register the distance between each pair of residues. Such distance is represented in a gray scale, with the maximum distance being 1.5 nm; if a distance between residues is larger than this value a black dot is added to this figure.

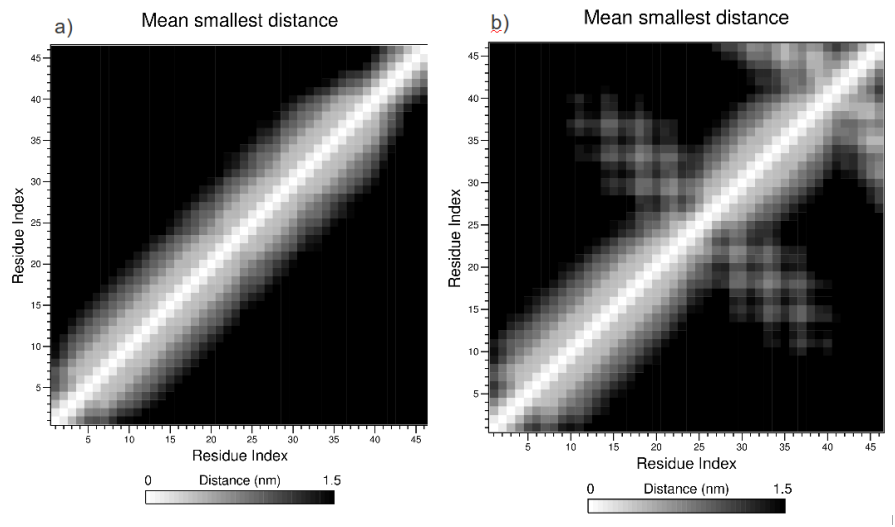

**Figure 4**. The mean smallest distance matrix for 0.5 M (a) and 0.6 M (b) concentrations of NaCl. The axes represent the residues that conform APOA1. The maximum distance considered to be part of the protein is 1.5 nm.

The principal diagonal line in Fig. 4 (in white) represents relative distance equal to zero, because it is the distance of a residue with itself. Therefore, a protein leaves a fingerprint of itself in a msdm graph. This is a useful tool to compare between the open and closed structures because it shows how the protein modifies its conformations from open to closed as it evolves in time and which residues are moving the most. In Fig. 4 we show two graphs, with

60                                                              M. A. Balderas Altamirano, et al.the left corresponding to an open APOA1 structure at a salt concentration equal to 0.5 M, and the right for the closed configuration at 0.6 M. The left graph (Fig. 4(a)) shows the msdm between the aminoacid residues that make up APOA1, where a diagonal line across the graph means that there are similar distances between the residues, which corresponds to the extended protein conformation (open). In Fig. 4(b), one observes qualitatively different behavior from the one found for the open conformation (left panel in Fig. 4). In the center of the graph on the right of Fig. 4, there is a gray area representing relative distances between residues that are smaller than 1.5 nm, meaning that the structure of the protein is that of the closed conformation and it also identifies the residues that are closer to each other because of this folding. We selected the residues 21 – 26 from the right panel in Fig. 4 because they are in the zone where the closed conformation is formed. These residues are H (Hystidine), L (Leucine), A (Alanine), P (Proline), Y (Tyrosine) and S (Serine). The sequence of APOA1 studied here is therefore called HLAPYS. In the upper right corner of the right panel in Fig. 4 we find another zone where residues are closer than 1.5 nm; these residues correspond to the end of the protein.

To determine which aminoacid residues are the responsible for the open and closed conformations discussed previously, we obtained the Ramachandran plot [Mathews, 2002], which maps out the conformations of the alpha carbon in a protein (or visualize backbone dihedral angles $\psi$ versus $\varphi$ of aminoacid residues in protein structure). In a Ramachandran plot one graphs the zone where the alpha carbon conformations are stable through time and also the zones where there are alpha helices and beta sheets. In Fig. 5 we show two graphs where the top (a) represents the open conformation (at 0.5 M of NaCl), and the bottom (b) corresponds to a closed structure, at 0.6 M. Black dots represent the configurations (over all the simulation time) of all the aminoacids in APOA1, while the different colors belong to specific aminoacid residues (HLAPYS) throughout the entire simulation time. In Fig. 5(a), one sees that the HLAPYS aminoacids are preferentially found in a relatively narrow zone of angle values: this is the region of the alpha helix zone. By contrast, in Fig. 5(b) we see that the HLAPYS residues move to the beta sheet area (top left quadrant in each graph), although this does not mean that they are forming beta sheets, instead it is just means that the aminoacids have values of their angles which are similar to those found in beta sheets. Also we can see that the aminoacids that having moving angles with time are PYS, while the other ones (H, L, A). Therefore the Ramachandran plot helps us identify specifically the residues that are responsible for the folding of the protein driven by the addition of ions.

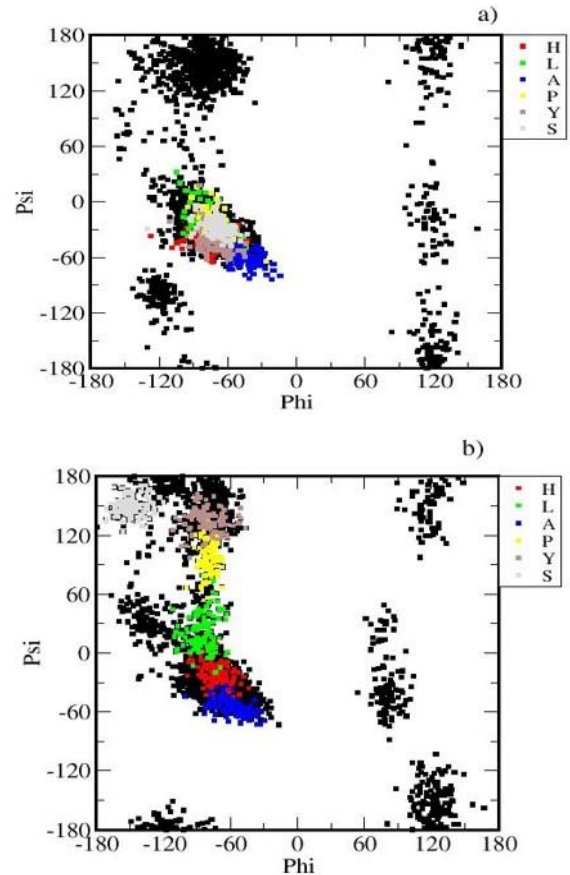

**Figure 5**. Ramachandran plot for the open configuration at 0.5 M (a), and for the closed one at 0.6 M of NaCl (b). The axes represent the angles that the $\alpha-C$ atom forms with its bonding neighbors. The colors and the symbols on the right captions represent the H, L, A, P, Y and S residues of the APOA1. The different data points for a given residue represent its evolution with time.

Up to this point we have studied the residues that drive APOA1 to form closed conformations.

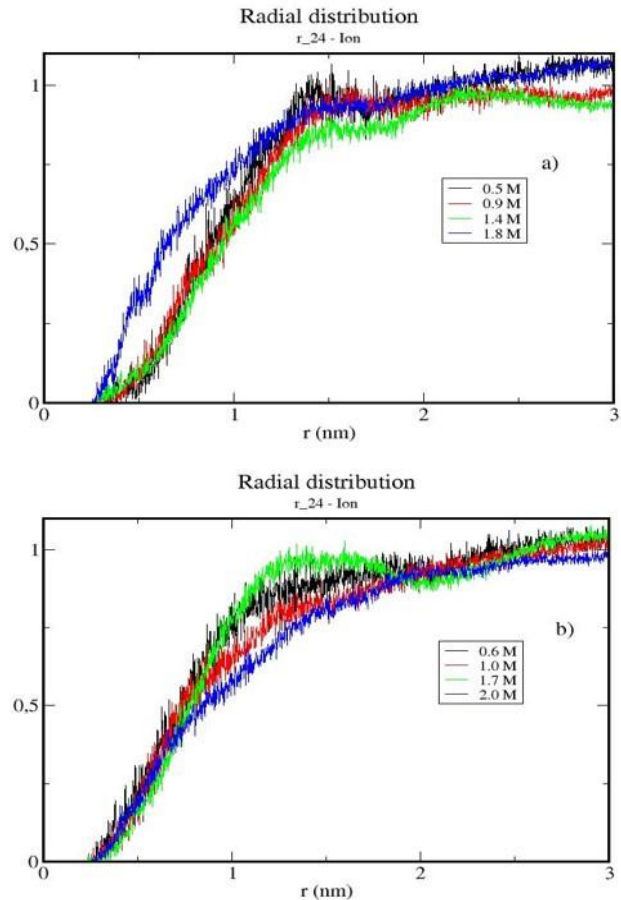

**Figure 6**. Averaged radial distribution function of the P residue with the both salt ions, at different NaCl concentrations, shown in various colors. (a) Open conformation; (b) closed conformation. The residue index 24 in the sequence of APOA1 is Proline (P).

Next we must understand the influence of the salt ions and water. With that purpose in mind we calculated the radial distribution function (rdf) of these residues and their interaction with water and with the ions. The rdf is obtained from the relative distance between a chosen particle in a fluid and all other particles that surround it; therefore it quantifies how the local density in a fluid varies [Allen, 1987]. In the case of water we did not see any significant difference between the rdf's of the water – open configuration, and those between the water – closed conformation, therefore we do not show them here. Instead we focused on calculating the rdf's between the ions and the residues. In Fig. 6 we show the rdf of the P – residue with the ions (namely, with the sum of Na and Cl ions).

In Fig. 6(a) we show the rdf of the P – residue and the ions (Na + Cl), at different concentrations for the case of an open conformation of APOA1. One feature that stands out is the relatively large number of ions in close proximity to the P residue at 1.8 M (blue line). Although P is a non – polar residue, it is highly movable, which implies that the ions can easily surround it and cause it to have a solvation radius larger than its own which in turn would compel the neighboring residues to stay at relatively larger distances from P. This fact, in turn, leads to an open configuration. In Fig. 6(b) there is a uniform distribution of relative distances between the P – residue and the salt ions at all concentrations; for example, at all salt concentrations in Fig. 6(b) there is approximately the same probability to have ions and the P – residue a distance $r \sim 0.7$ nm apart. The fact that all rdf's in this case follow the same trend at small relative distances means that these rdf's correspond to configurations where there is no more room available for the ions to get closer to P. This must be the case when APOA1 is in a closed conformation, as in Fig. 6(b).

## 4 Conclusions

We have shown in this work how atomistically detailed molecular dynamics computer simulations can help us understand the folding process in proteins under controlled physicochemical conditions as is, for example, the ionic strength. For this case study we chose APOA1 because of its key role in producing the efflux of fatty acids in the human blood stream. It was found that, under increasing salt concentration, the protein undergoes alternatively folding and unfolding, which the simulations suggest that it is driven by the formation of solvation spheres around the most mobile aminoacids in the APOA1 sequence, competing with the electrostatic interactions. The structural properties of the protein were characterized through powerful tools such as the radius of gyration, the end to end distance, the mean square distance matrix, the Ramachandran plot, and the radial distribution function. In particular, we found that the P, Y and S residues are moving the most when going from an open to a close conformation. This work helps elucidate the effect of the charge in APOA1, where the important sequence turns out to be HLAPYS. The role of the closed configurations should not be underestimated either, for those are precisely the ones thought to be responsible for the trapping of fatty acids such as cholesterol. Work is presently under way to determine the association of APOA1 with cholesterol under varying ionic strength, but it is indispensable to gain first a basic understanding of this protein's folding at increasing salt content before attempting the study of its complexation with other molecules.

## Acknowledgements

This project was financed by CONACYT, grant 132056. We would like to acknowledge many educational conversations on this topic with A. López – Vallejo López. The authors thank also J. Limón (IFUASLP) for technical support with GPU installation. MABA and AGG thank the UASLP for its hospitality.